\documentclass[a4paper,11pt]{article}
\usepackage{aaskaiid}
\usepackage{orcidlink}
\usepackage{amsmath, bm, amssymb, dsfont}

\usepackage{multirow, makecell}
\usepackage{subcaption, graphicx}

\title{Weak Lensing with the SKAO: Radio Shear Measurement}
\ShortTitle{Radio Shear Measurement}

\author[1]{Priyamvad Tripathi\orcidlink{0000-0003-0294-2512}}
\ShortName{Tripathi et al.} 
\author[2]{Marzia Rivi\orcidlink{0000-0002-2549-7813}}
\author[1]{Simon Prunet\orcidlink{0000-0002-1755-4582}}
\author[3]{Ian Harrison\orcidlink{0000-0002-4437-0770}}
\author[1]{Andr\'e Ferrari\orcidlink{0000-0002-7297-7625}}

\affiliation[1]{Université Côte d’Azur, Observatoire de la Côte d’Azur, CNRS, 06000 Nice, France}
\emailAdd{priyamvad.tripathi@oca.eu}
\affiliation[2]{Dopartimento di Ingegneria, Università di Modena e Reggio Emilia, Modena, Italy}
\emailAdd{marzia.rivi@unimore.it}
\affiliation[3]{School of Physics and Astronomy, Cardiff University, The Parade, Cardiff, Wales, UK CF24 3AA}

\abstract{Cosmic shear measurements have traditionally been dominated by optical surveys, which offer higher resolution, better sensitivity, higher galaxy number density, and wider area coverage than their radio counterparts. With the advent of upcoming radio surveys, particularly those planned with the SKA-Mid, we would reach the high sensitivity and resolution required for weak lensing studies, allowing radio observations to begin competing with optical surveys in this domain. However, radio observations are affected by fundamentally different instrumental and astrophysical systematics, meaning that shape and shear measurement techniques developed for optical surveys cannot be straightforwardly applied. Fully realizing the weak lensing potential of these next generation radio surveys therefore requires the development and validation of methods tailored specifically for radio datasets.\\

In this chapter, we present a simulation pipeline to generate realistic observations of isolated radio galaxies based on the SKA-Mid AA4 array configuration. We apply three recent radio shape measurement techniques—SuperCALS, RadioLensfit, and DeepShape—to assess their performance on these simulations. Our results show that RadioLensfit and DeepShape yield the most accurate shape estimates, although RadioLensfit is significantly more computationally intensive. We also perform shear recovery on simulated data using RadioLensfit and DeepShape, finding multiplicative and additive shear biases of order a few $10^{-2}$ and $10^{-4}$, respectively. Finally, we highlight the challenge of source separation, which will play a critical role in the success of future radio weak lensing analyses.
}

\begin{document}
\maketitle

\section{Introduction}

Cosmic shear is a powerful tool for probing the total mass distribution and studying the relationship between dark matter and baryonic matter. Since shear only contributes a small distortion to the intrinsic galaxy shape, extracting the shear signal requires both highly precise shape measurements and a large sample of galaxies. To date, most cosmic shear detections have been obtained in the optical waveband~\citep{optical-1,optical-2,optical-3,optical-4}, primarily due to the higher angular resolution, superior sensitivity, greater background galaxy densities, and wider survey coverage achievable at optical wavelengths. Nevertheless, cosmic shear has also been detected in the radio band, albeit with lower statistical significance~\citep{Chang_2004}, as well as through cross-correlations with optical galaxies~\citep{cross-1,cross-2}. These early detections demonstrate the feasibility of radio weak lensing, but also highlight the need for accurate and scalable shape measurement techniques tailored to radio observations.

Shape measurement techniques are highly advanced in the optical domain, largely because most weak lensing studies have historically been conducted at optical wavelengths, which provide high sensitivity, angular resolution, and large galaxy number densities. To fully realise the potential of weak lensing with the SKA, it is necessary to assess whether optical methods can be directly applied to radio observations, as well as to develop new techniques tailored to radio data. Unlike optical telescopes, radio interferometers measure visibilities, which correspond to discrete samples in the Fourier domain of the sky brightness distribution. Early work therefore focused on performing shape measurements directly in the visibility domain \citep{Chang_2004, Patel_2010}, typically by fitting parametric galaxy models to the observed visibilities \citep{Priti_Patel, Rivi, Rivi3, Rivi2}. Working directly with visibilities avoids biases introduced during the imaging and deconvolution process. However, the reliance on parametric models may limit the ability to capture the full diversity of galaxy morphologies, and the large volume of visibility data makes these methods computationally expensive and difficult to scale to future surveys such as SKA.

To improve scalability, alternative approaches perform shape measurement in the image domain. The visibilities are inverted to form a dirty image, followed by deconvolution using methods such as CLEAN or its variants \citep{Clean} to remove the effects of PSF. Additional corrections, such as those for the $w$-term and primary beam, are typically applied during this stage to further improve image fidelity. Once a sufficiently accurate estimate of the sky is obtained, cutouts of individual sources are extracted, and their shapes measured using moment-based methods \citep{galsim} or parametric model fitting \citep{im3shape, Pybdsf}. This approach is significantly faster and scales better to large datasets, but the accuracy is limited by the fidelity of the image reconstruction, and deconvolution errors could propagate into biased shape estimates \citep{Priti_Patel}. To address this, recent surveys such as \citet{Superclass} apply simulation-based bias corrections to the original shape estimate for better accuracy. More recently, deep learning methods have emerged as a complementary approach, combining learned image reconstruction with dedicated shape measurement networks to improve accuracy and robustness~\citep{DeepShape}.

Motivated by these developments and the need for robust validation frameworks, in this chapter we introduce a simulation pipeline for generating realistic radio observations that can serve as testbeds for shear measurement methods. We then highlight and evaluate three representative approaches for estimating galaxy shapes and shear from these simulations. Finally, we discuss the critical challenge of source separation, which poses a major obstacle for future radio weak lensing analyses.

\section{Simulation Pipeline}
\label{sec:simulation}
We simulate realistic SKA-Mid observations of isolated star-forming galaxies (SFGs) using a pipeline based on the Tiered Radio Extragalactic Continuum Simulation (T-RECS)~\citep{TRECS}. A full description of the pipeline can be found in~\citet{DeepShape}; here, we summarize the key steps. In this work, we restrict our analysis to isolated radio galaxies for simplicity, ensuring no contamination from neighbouring sources in either the image or visibility domain. This represents an idealized best-case scenario, as in real observations such blending and contamination can significantly bias shape measurements and must be carefully mitigated.

\subsection{Galaxy Population}
We use the 400 deg$^2$ SFG-wide catalog from T-RECS, which provides source flux densities ($I_0$ at $\nu_0 = 1.4$ GHz), sizes, ellipticities, and sky positions. We restrict our sample to galaxies with {$50~\mu$Jy $\leq I_0 < 200~\mu$Jy} to ensure uniform dynamic range and exclude bright sources less relevant for weak lensing.

Each galaxy is modeled with a Sérsic brightness profile:
\begin{equation}
    I(r) = I_e~ \text{exp} ~\left\{-b_{s}\left[\left(\frac{r}{r_e}\right)^{1/s}-1\right]\right\},
\end{equation}
where $I_e$ is the flux at effective radius $r_e$, $s \sim \mathcal{U}(0.7, 2)$ is the Sérsic index, and $b_{s}$ is a scaling factor~\citep{Sersic}. All galaxy images are simulated on a $128 \times 128$ grid with a pixel scale of $\theta_{\text{pix}} = 0.143^{\prime\prime}$ using the GalSim package~\citep{galsim}. The pixel intensities are normalized such that the total flux of each galaxy image equals $I_0$. The two quantities are related as ${I_e = \frac{I_0}{2\pi s r_e^2}.\frac{b_s^{2s}}{e^{b_s}\Gamma(2s)}}$, where $\Gamma$ is the gamma function.
\subsection{Interferometer Model}
We compute visibilities using the SKA-Mid AA4 configuration with a central frequency of $\nu_0=1.4$ GHz and a fractional bandwidth of $\Delta\nu=0.3\nu_0$ (Band 2). Observations span 8 hours with integration time per visibility $\tau_{\mathrm{int}}=300$s, resulting in 1,853,376 visibilities for each source.

\begin{figure}[ht]
   \centering
\subfloat{
           \includegraphics[height=6.2cm, width=6.2cm, keepaspectratio]{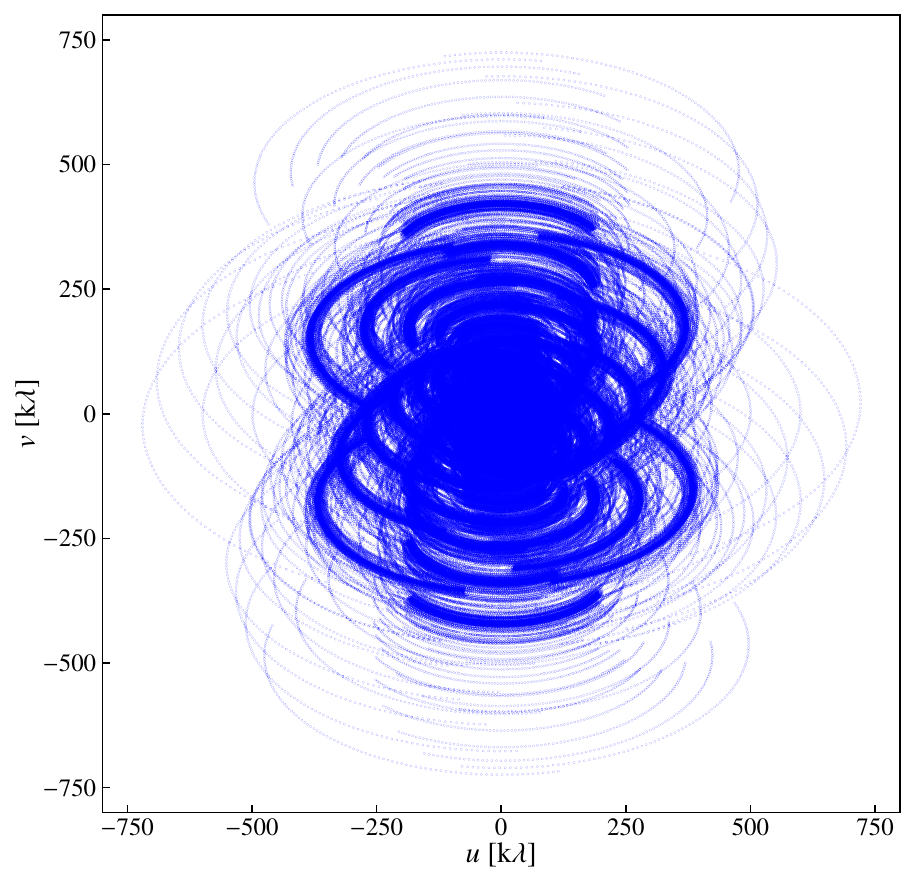}
        }
        \hfill
        \subfloat{
            {\includegraphics[height=6.2cm, width=6.2cm, keepaspectratio]{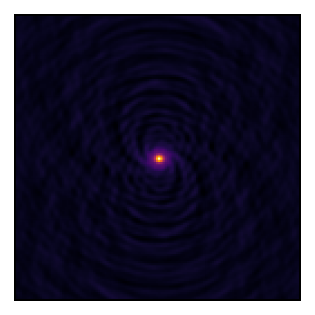}}
        }   
      \caption{\textit{Left:} Visibility coverage for a pointing at declination of {$\delta=-30^\circ$}. $u$ and $v$ are plotted in 1000 wavelength units k$\lambda$. Reproduced from \citet[Fig.~3]{DeepShape}. \textit{Right:} The corresponding PSF using robustness $R=-0.5$.}
         \label{fig:uv}
   \end{figure}

Radio surveys map the sky through multiple pointings, each covering a field of view with several galaxies. As a result, each galaxy has a unique effective PSF depending both on the pointing direction and the galaxy’s position relative to it. To approximate this effect, we centered each pointing on the target galaxy, thereby assigning a unique PSF to every source. While this simplification ignores intra–pointing PSF variations, such effects are expected to be negligible at mid-frequencies where the primary beam spans $\sim1~\mathrm{deg}^2$ \citep{Atemkeng}. Fig.~\ref{fig:uv} illustrates the visibility coverage ($uv$ coverage) for a pointing at declination of $\delta=-30^\circ$ and the corresponding PSF.

Gaussian noise is added to each visibility, with RMS noise $\sigma_V$ calculated as:
\begin{equation}
    \sigma_V = \frac{2k_BT_{\text{sys}}}{\eta A_{\text{eff}}} \times \frac{1}{\sqrt{2\tau_{\text{int}}\Delta\nu}},
\end{equation}
where all system parameters are based on SKA-Mid specifications~\citep{SKA_Details}, resulting in $\sigma_V \approx 0.5$ mJy. Simulations were performed using the Radio Astronomy Simulation, Calibration, and Imaging Library (RASCIL) package\footnote{\href{https://ska-telescope.gitlab.io/external/rascil/}{RASCIL Documentation}}.

During imaging, we weight the visibilities according to the  Briggs (robust) weighting scheme~\citep{Robust}. This scheme (approximately) interpolates between natural and uniform weighting based on a tunable robustness parameter $R$~\citep{CASA}. After testing different values, we adopt $R = -0.5$ as an effective compromise between resolution and sensitivity\footnote{see \citet[Appendix B]{DeepShape} for discussion}, resulting in an RMS pixel noise of $\sigma \approx 0.71~\mu$Jy.

\section{Shape Measurement}

\subsection{Methods}

In this work, we test three representative radio shape measurement approaches spanning visibility based, image based, and deep learning methodologies. This section describes each method in detail, along with the modifications made to apply them to our simulated datasets.
\subsubsection{SuperCALS}
SuperCALS is a shape measurement method that operates in the image domain~\citep{Superclass}, and was recently used in the SuperCLASS radio weak-lensing survey~\citep{Superclass1}. It is a two-stage algorithm involving \textit{source-level} and \textit{population-level} calibration.

In the \textit{source-level} calibration, the initial shape estimate $\bm{\epsilon}^{\text{est}}$ of a galaxy is obtained from the multiscale CLEAN (MS-CLEAN)~\citep{MSCLEAN} deconvolved image. To quantify measurement bias for this source, model galaxies with known ellipticities $\bm{\epsilon}^{\text{inp}}$ are injected into the residual MS-CLEAN image. The observed ellipticities $\bm{\epsilon}^{\text{obs}}$ are then used to estimate the source-specific bias $\bm{b}^S \in \mathds{R}^2$ as:
\begin{equation}
    \bm{b}^S(\bm{\epsilon}^{\text{inp}}) \equiv \bm{\epsilon}^{\text{obs}} - \bm{\epsilon}^{\text{inp}}.
\end{equation}
This bias is modeled as a second-order 2D polynomial in $\bm{\epsilon}^{\text{inp}}$. The initial ellipticity estimate is then corrected using this model:
\begin{equation}
    \hat{\bm{\epsilon}} = \bm{\epsilon}^{\text{est}} - \bm{b}^S(\bm{\epsilon}^{\text{est}}).
\end{equation}

Despite this correction, residual bias often remains. A second \textit{population-level} calibration step is applied using a simulated galaxy ensemble to model and correct the remaining bias in the source-calibrated shapes.

We use the original SuperCALS implementation\footnote{\href{https://github.com/itrharrison/supercals/}{https://github.com/itrharrison/supercals/}} with minor adaptations to accommodate our simulated dataset. Specifically, we use the HSM module from GalSim~\citep{galsim} for shape measurement in place of Im3Shape~\citep{im3shape}. In addition, the ellipticities of the model sources are restricted to $\epsilon_\alpha = \{0, \pm0.1375, \pm0.275, \pm0.4125, \pm0.55\}$ where $\alpha \in {1,2}$ denotes the two ellipticity components. This range reflects the ellipticity distribution for the galaxies in our test set and also avoids slow convergence issues encountered when using HSM shape measurement on highly elliptical sources.
\subsubsection{RadioLensfit}
RadioLensfit is a model-fitting method that operates directly in the visibility domain, thus avoiding any imaging-related biases \citep{Rivi}. It assumes galaxies follow an exponential (Sérsic index $n_s=1$) brightness profile. Under this model, the predicted visibilities are:
\begin{equation}
    \bm{v}_m(u,v,w)= \frac{I_0\, e^{2\pi j(ul+vm+w(n-1))}}{(1+4\pi r_\alpha^2|\Tilde{\bm{A}}^{-T}\bm{k}|)^{3/2}},
\end{equation}
where $\bm{k} = (u,v)$, $I_0$ is the flux, $r_\alpha$ is the scale radius, and $\Tilde{\bm{A}}$ encodes the ellipticity:
\begin{equation}
    \Tilde{\bm{A}}=
    \begin{bmatrix}
        1 - \epsilon_1 & -\epsilon_2\\
        -\epsilon_2 & 1 + \epsilon_1
    \end{bmatrix}.
\end{equation}
Assuming Gaussian noise, the likelihood for parameters $\bm{\Theta} = \{I_0, \alpha, l, m, \epsilon_1, \epsilon_2\}$ is:
\begin{equation}
    p(\bm{v}_d|\bm{\Theta}) \propto \exp\left[-\frac{(\bm{v}_d - \bm{v}_m)^\dagger \bm{C}^{-1}(\bm{v}_d - \bm{v}_m)}{2}\right].
\end{equation}
To reduce the data volume, which would be very expensive both in terms of memory size and computational time, data visibilities are averaged in a coarse regular uv grid, and model visibilities are directly evaluated for the uv points of this grid.   
This likelihood is marginalized over nuisance parameters— $\{I_0,\alpha,l,m\}$— using appropriate priors. The ellipticity is then estimated by adaptively sampling the marginalized likelihood. We guide the reader to~\citet{Rivi2} for details on prior setting and likelihood sampling. 

We use the publicly available RadioLensfit implementation\footnote{\href{https://github.com/marziarivi/RadioLensfit2/}{https://github.com/marziarivi/RadioLensfit2/}}~\citep{Radiolensfit_code} running in OpenMP mode for computational efficiency. RadioLensfit requires specifying prior parameters for flux and size, as well as the visibility noise. The default values are based on the VLA 20 cm continuum catalogue; we modify them to match the distributions and noise levels used in our simulations. Details of the modified parameters are given in \citet{DeepShape}.

\subsubsection{DeepShape}
DeepShape is a fully supervised deep learning framework that predicts the ellipticity of a galaxy, $\hat{\bm{\epsilon}}$, from a given dirty image of an isolated radio galaxy, and the associated PSF. DeepShape is made of two modules:
\begin{enumerate}
\item {The first module is an image reconstruction module based on a Plug-and-Play (PnP) approach. This module iteratively deconvolves the PSF and removes noise from the dirty image to produce an estimate of the true galaxy.}
\item {The second module is a shape measurement network that estimates the galaxy shape from the reconstructed image while explicitly accounting for residual PSF effects, ensuring that the predicted ellipticity is robust to any remaining PSF distortions.}
\end{enumerate}

\paragraph{Image reconstruction}  
The reconstruction problem is formulated as a regularized deconvolution, aiming to recover the true galaxy image while enforcing a prior via the denoiser. This leads to the following optimization problem:
\begin{equation}
\label{eq:decon}
\hat{\bm{i}}=\arg\min_{\bm{i}}\; \frac{1}{2\sigma^2}\|\bm{i}_D - H\bm{i}\|_2^2 + \lambda \Phi(\bm{i}),
\end{equation}

where $H$ denotes convolution with the PSF kernel $\bm{h}$, $\sigma$ is the RMS noise level, $\Phi$ is a regularization function encoding the log-prior on the true image, and $\lambda > 0$ is the regularization weight.  

To solve Eq.~\ref{eq:decon}, an auxiliary variable $\bm{z}$ is introduced and the problem is addressed via Half-Quadratic Splitting with alternating updates:
\begin{align}
\bm{i}_{k+1} &= \big(H^\dagger H + \mu_k \sigma^2 \bm{I}\big)^{-1} \big(H^\dagger \bm{i}_D + \mu_k \sigma^2 \bm{z}_k\big), \\
\bm{z}_{k+1} &= \mathrm{DN}_{\hat{\theta}}\big(\bm{i}_{k+1};\, \sigma_k\big), \qquad \sigma_k \equiv \sqrt{\lambda / \mu_k},
\end{align}
where $\mathrm{DN}_{\hat{\theta}}$ denotes a learned denoiser. A DRUNet network \citep{ZhangPnP} serves as the denoiser, which is trained on a dataset of 250\,000 radio galaxies simulated using the same pipeline. This enables the module to encode a strong prior for isolated radio galaxies, effectively removing noise and deconvolving the PSF while preserving intrinsic galaxy morphology.

\paragraph{Measurement network}  
Following reconstruction, the shape measurement network predicts galaxy ellipticity from the reconstructed image and its corresponding PSF. The network processes the inputs through two parallel branches to separately capture intrinsic galaxy features and residual PSF effects:

\begin{itemize}
    \item \textbf{Galaxy feature extraction} – An $\mathrm{E}(2)$-equivariant CNN block $\mathcal{F}$ extracts a feature vector $\bm{f} = \mathcal{F}(\hat{\bm{\imath}})$ from the reconstructed image. The equivariance property ensures that the extracted features respond consistently to image rotations, translations, and reflections, preserving the geometric information needed for accurate shape measurement.
    
    \item \textbf{PSF encoding} – A pre-trained encoder $\mathcal{E}$ maps the PSF to a low-dimensional representation $\boldsymbol{\zeta} = \mathcal{E}(\bm{h})$, enabling the network to model and correct residual PSF effects. The autoencoder is trained on a simulated dataset of 100,000 SKA-Mid PSFs.
\end{itemize}

The vectors $\bm{f}$ and $\boldsymbol{\zeta}$ are concatenated and passed to a dense layer $M_{\theta}$, which outputs the predicted ellipticity: $\hat{\boldsymbol{\epsilon}} = M_{\theta}([\bm{f}, \boldsymbol{\zeta}])$. The shape measurement network is trained in two stages using datasets simulated with the same pipeline: first on 250\,000 noiseless galaxy images, and then fine-tuned on 250\,000 reconstructed galaxy–PSF pairs. All datasets for training stages are generated using the same simulation pipeline, and an independent test set produced by the same pipeline is used to evaluate DeepShape and other methods in the following section. We refer the reader to \cite{DeepShape} for details on the network structure and hyperparameter optimization.

\subsection{Results}

\begin{figure}[ht]
    \centering
	\includegraphics[width=0.75\columnwidth]{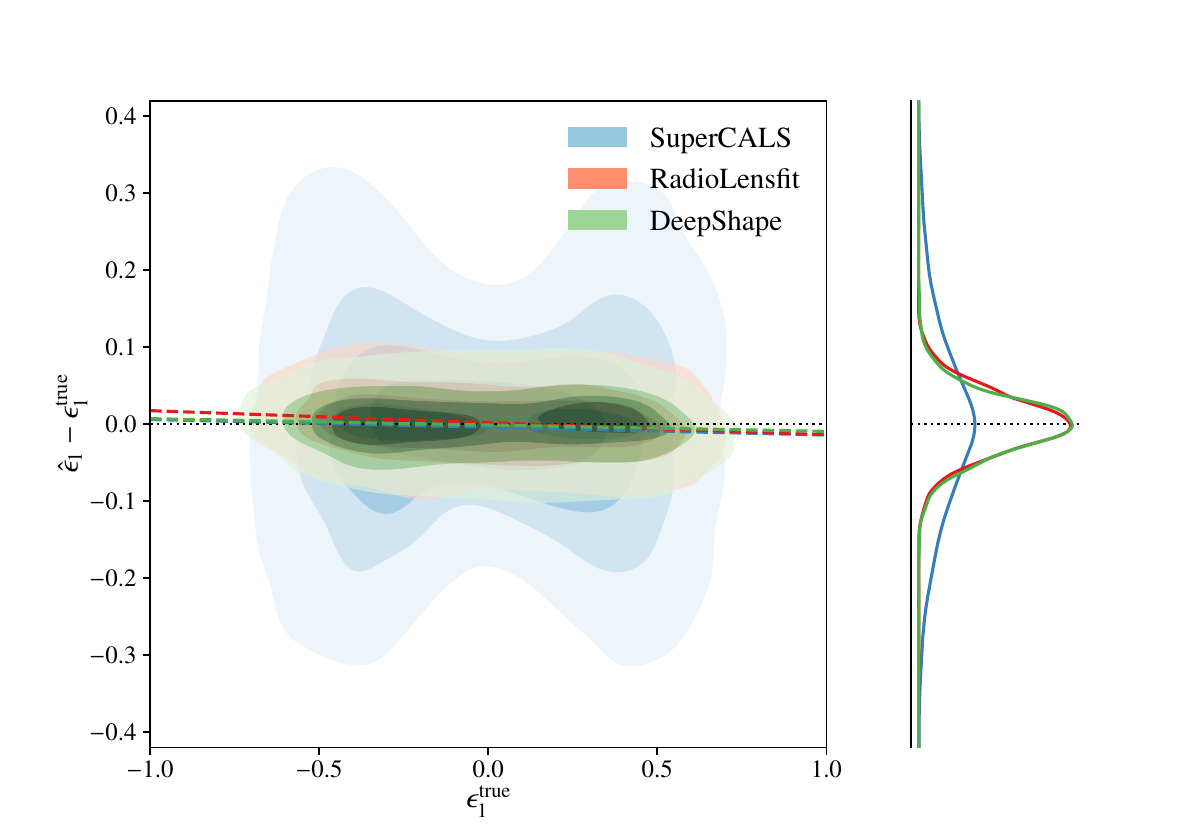}
    \caption{Residuals of the first ellipticity component ($\epsilon_1$) for different shape measurement methods as a function of the input value. The 2D contours are obtained via kernel density estimation, with colored dashed lines indicating the best-fit linear relationship. The right panel shows the corresponding marginal 1D distribution of the residuals.}
    \label{fig:shape_plot}
\end{figure}

We assess the performance of the methods using a simulated test set containing $2\,500$ galaxies with varying sizes, fluxes, shapes, and PSFs. To quantify prediction accuracy, we compute the root mean square error (RMSE) for each ellipticity component as:
\begin{equation}
\label{eq:rmse}
\mathrm{RMSE}_\alpha = \sqrt{\langle (\hat{\epsilon}_\alpha - \epsilon_\alpha^{\mathrm{true}})^2 \rangle},
\end{equation}
where $\alpha \in (1, 2)$, $\epsilon_\alpha^{\mathrm{true}}$ is the true ellipticity, and $\hat{\epsilon}_\alpha$ is the predicted value. We also evaluate the Pearson correlation coefficient $\rho_\alpha$ for each ellipticity component:
\begin{equation}
\rho_\alpha = \frac{\mathrm{cov}(\hat{\epsilon}_\alpha, \epsilon_\alpha^{\mathrm{true}})}{\sigma(\hat{\epsilon}_\alpha) \sigma(\epsilon_\alpha^{\mathrm{true}})},
\end{equation}
where $\mathrm{cov}(\hat{\epsilon}_\alpha, \epsilon_\alpha^{\mathrm{true}})$ denotes the covariance between predictions and ground truth, and $\sigma(\cdot)$ indicates the standard deviation. Additionally, we quantify shape measurement bias by fitting a linear model to the relationship between the true ellipticity $\epsilon_\alpha^{\mathrm{true}}$ and the residuals $(\hat{\epsilon}_\alpha - \epsilon_\alpha^{\mathrm{true}})$, from which we estimate the multiplicative bias $\hat{m}_\alpha$ (the slope) and the additive bias $\hat{c}_\alpha$ (the intercept). A summary of the comparison metrics for the first ellipticity component is provided in Table~\ref{tab:shape}. The final two columns show the processing time per source for each method and the fraction of sources successfully resolved in the test set, respectively. The residuals for the first ellipticity component are shown in Fig.~\ref{fig:shape_plot}.

All methods achieve similar performance in terms of linear bias. SuperCALS, however, shows noticeably higher variance, as reflected in its RMSE, Pearson coefficient, and larger uncertainties in the bias parameters. This likely stems from MS-CLEAN’s insufficient reconstruction accuracy for weak lensing, as previously demonstrated \citep{Priti_Patel, Polish, DeepShape}. In terms of inference speed, SuperCALS is the fastest, followed closely by DeepShape, whereas RadioLensfit is significantly slower because it operates directly on the full set of visibilities, which is typically very large. It should be noted that these times reflect only inference: DeepShape requires additional training time, while SuperCALS requires time to generate simulated datasets for \textit{population level} calibration. RadioLensfit also fails to resolve a notable fraction of sources, due to highly noisy likelihood evaluations.

Since our analysis is limited to isolated galaxies, the performance of all three methods represents an upper limit to their achievable accuracy. In wide-field observations, however, additional complications arise from source crowding and blending. In such cases, SuperCALS and DeepShape rely on external tools such as PyBDSF for source extraction and separation \citep{Pybdsf}, whereas RadioLensfit employs a faceting technique to extract source visibilities \citep{Rivi_multi}.
\begin{table}[ht]
\centering
\caption{Performance comparison of shape measurement methods for ellipticity component $\epsilon_1$, showing evaluation metrics, time per source, and percentage of sources resolved. Only $\epsilon_1$ results are displayed for brevity. The best metric in each column is highlighted in bold.}
\label{tab:shape}
\begin{tabular}{c||c|c|c|c|c|c}
\multirow{2}{*}{Method} & \multirow{2}{*}{$\text{RMSE}_1$} & \multirow{2}{*}{$\rho_{1}$} & {$\hat{m}_1$} & {$\hat{c}_1$} & {Time} & {$\%$ Resolved}\\
{} & {} & {} & {$[10^{-3}]$} & {$[10^{-4}]$} & {$[s]$} & {} \\
\hline
\hline
{SuperCALS} & $0.122$ & $0.950$ & {$-10.2\pm7.2$} & {$-39.6\pm24.4$} & $\bm{0.18}$ & {$98.49$} \\
{RadioLensfit} & $\bm{0.040}$ & $\bm{0.992}$ &$-15.7\pm2.9$&$19.3\pm9.5$&$230$&$72.16$\\
{DeepShape} & ${0.041}$ & ${0.990}$ & {$\bm{-8.3\pm2.5}$} & {$\bm{-10.7\pm8.3}$} & ${0.22}$ & {$\bm{100}$}\\
\end{tabular}
\end{table}
\section{Shear Measurement}
Weak lensing studies depend on the accurate recovery of shear from intrinsically noisy galaxy shape measurements, making the choice of shape estimation method critical. To explore the effectiveness of current techniques, we examine two complementary approaches: an image-domain analysis using DeepShape and a visibility-domain analysis using RadioLensfit. Both methods are tested on realistic SKA-Mid simulations, and their recovered linear shear biases are benchmarked against the requirements for weak lensing surveys with SKA-Mid. The accuracy of measured shear is defined using a linear bias model as \(\gamma^\mathrm{meas}_\alpha = (1 + M_\alpha) \gamma^\mathrm{true}_\alpha + C_\alpha\), 
where $M_\alpha$ and $C_\alpha$ denote the multiplicative and additive biases for each shear component $\alpha\in\{1,2\}$. For a survey spanning 5000~deg$^2$, the systematic error budget imposes limits of $|M_\alpha|\leq6.7\times10^{-3}$ and $|C_\alpha|\leq8.2\times10^{-4}$ \citep{brown2015}, ensuring that systematic errors remain subdominant to statistical uncertainties. In this analysis, we aim to determine whether these constraints can be achieved with existing methods. The image- and visibility-domain analyses presented in Secs.~4.1 and 4.2 use independent simulated datasets generated with different, but comparable, SKA-Mid simulation pipelines. These datasets were developed in prior studies to evaluate each method under realistic observing conditions. While not identical, both are representative of SKA-Mid observations and allow meaningful assessment of shear recovery performance.

\subsection{Image domain}
We first assess shear recovery in the image domain using DeepShape on a test set of one million galaxies simulated using the pipeline described in Section~\ref{sec:simulation}. The test set is split into 100 shear fields of 10,000 galaxies each. The two shear components, $\gamma_\alpha$, were sampled from a uniform distribution $\mathcal{U}(-0.1,0.1)$. Within each field, galaxies varied in flux, size, and intrinsic shape, but shared the same common PSF. To suppress shape noise, galaxies were simulated in $90^\circ$ rotated pairs, ensuring the mean intrinsic ellipticity per field was zero. 

\begin{figure}[ht]
\centering
\includegraphics[width=0.47\textwidth]{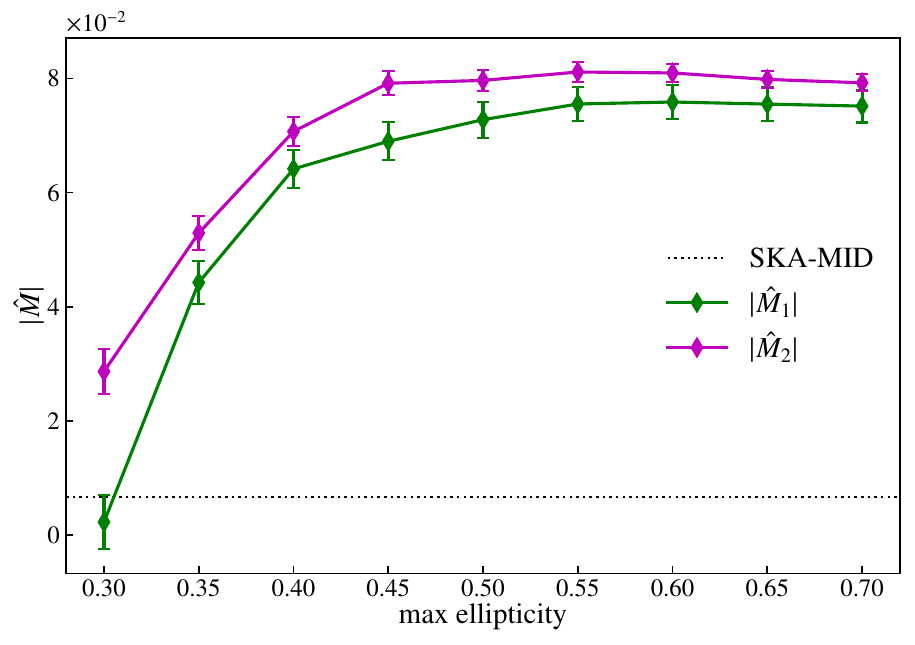}
\hfill
\includegraphics[width=0.47\textwidth]{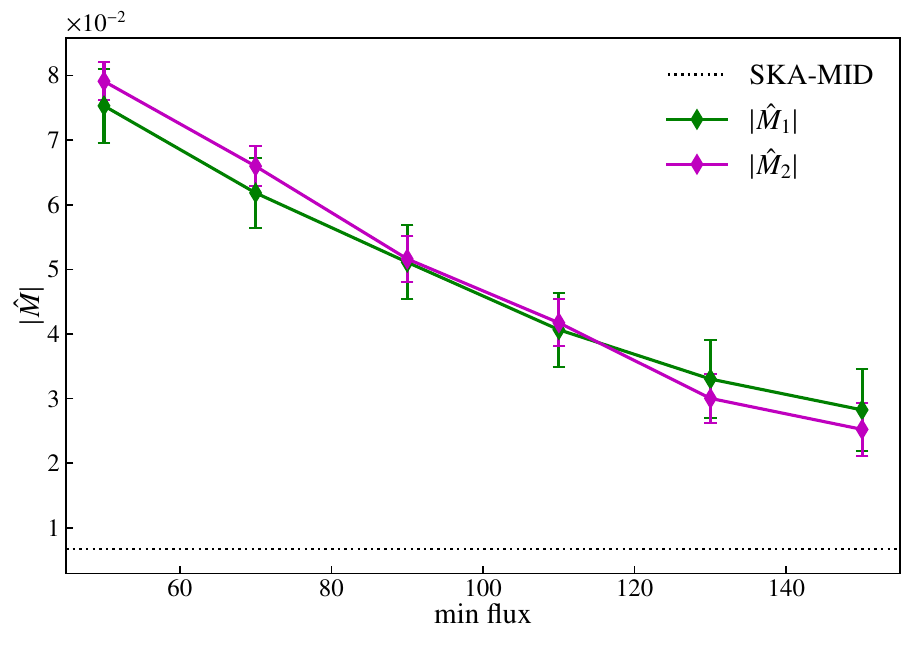}
\caption{Estimated multiplicative shear bias ($\hat{M}_1$, $\hat{M}_2$) for the two shear components as a function of galaxy selection cuts, obtained from the slope of the best fit linear relation between the measurement residuals and the true shear. The dotted black line indicates the SKA Mid requirement. \textit{Left}: dependence on maximum ellipticity. \textit{Right}: dependence on minimum flux. Reproduced from~\citet[Fig.~10]{DeepShape}.} 
\label{fig:m_depend}
\end{figure}

For each field, shear was estimated as the ensemble average of ellipticities predicted by DeepShape. Multiplicative and additive biases, $\hat{M}_\alpha$ and $\hat{C}_\alpha$, were obtained from the best-fit linear relation between the shear residuals and the true shear. The resulting bias estimates for the two components are:
\begin{equation*}
\begin{aligned}
    \hat{M}_1&=(-75.3\pm5.7)\times10^{-3},~\hat{C}_1=(-1.9 \pm 2.5)\times10^{-4}; \\
    \hat{M}_2&=(-79.1\pm2.9)\times10^{-3},~\hat{C}_2=(3.7 \pm 1.2)\times10^{-4}.
\end{aligned}
\end{equation*}

While the additive bias is close to the SKA-Mid requirements, the multiplicative bias is roughly an order of magnitude too high. However, it shows significant dependence on galaxy selection. In particular, applying a maximum ellipticity cut or a minimum flux cut reduces $\hat{M}_\alpha$ to levels closer to the required limit. This dependence is illustrated in Fig.~\ref{fig:m_depend}.  
\subsection{Visibility domain}
We also perform a complementary analysis in the visibility domain using RadioLensfit. The dataset used for this study was simulated using a visibility-domain simulation pipeline~\citep{Rivi}, which differs from the image-domain simulation pipeline described in Section~\ref{sec:simulation} but uses similar galaxy population models and observational parameters. Shear estimates were obtained for eight fields with uniformly oriented input shear values of amplitude $|\bm{\gamma}| = 0.04$, as well as for $|\bm{\gamma}| = 0$. Each field contained 10,000 simulated galaxies with fluxes in the range $10~\mu$Jy $\leq I_0 < 200~\mu$Jy, corresponding to a signal-to-noise ratio, SNR $\geq 10$. The visibility-domain SNR is defined as $\textrm{SNR} = \sqrt{\sum_{i=1}^\textrm{nvis} |\bm{v}_i|^2/\sigma^2_i}$, where $\bm{v}_i$ are the noiseless visibilities and $\sigma_i$ is the standard deviation of the visibility noise. Fluxes, sizes, and ellipticities were drawn from realistic distributions, and ten equally spaced orientations were assigned to each ellipticity modulus to suppress shape noise.

\begin{figure}[ht]
\centering
\includegraphics[width=0.47\textwidth]{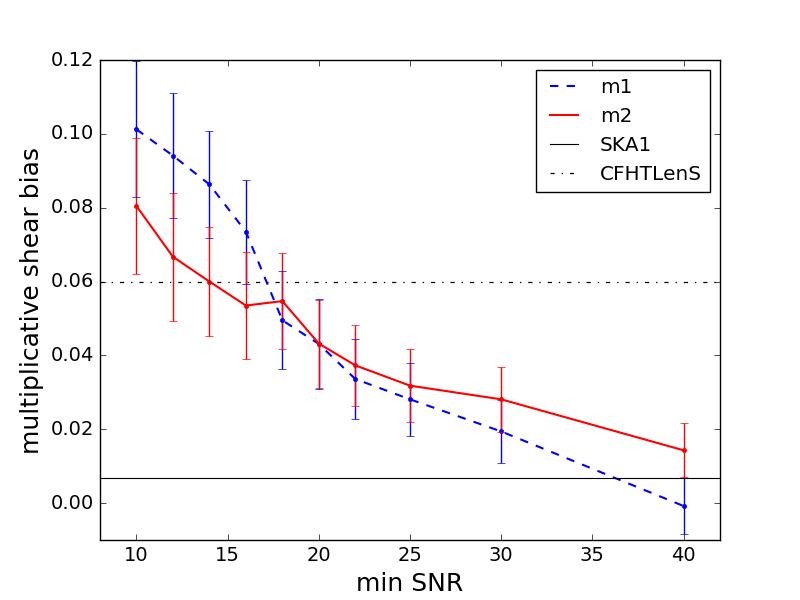}
\hfill
\includegraphics[width=0.47\textwidth]{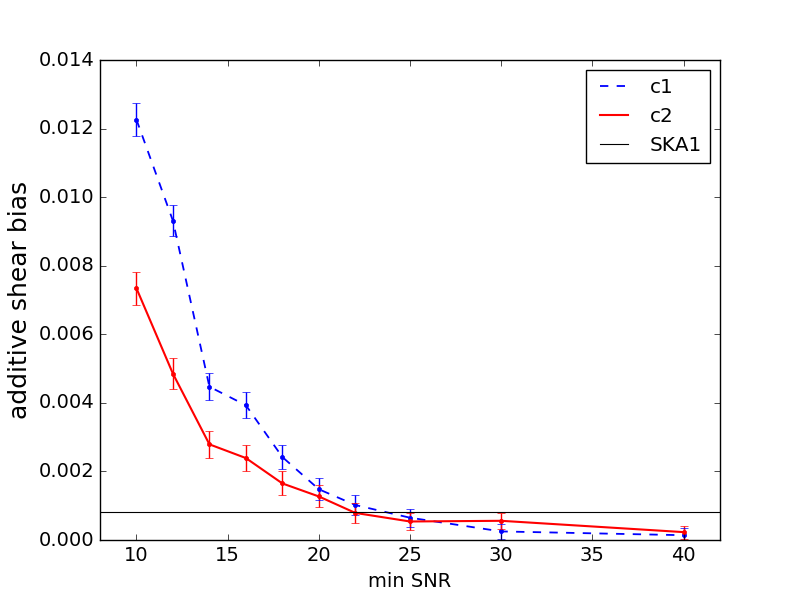}
\caption{\textit{Left}: Multiplicative shear bias for both components as a function of minimum SNR, compared with SKA-Mid requirements (solid black line) and CFHTLenS calibration correction (dash-dotted line) \citep{Heymans2012}. \textit{Right}: Additive shear bias for both components as a function of minimum SNR. Reproduced from~\citet[Figs.~7 and 8]{Rivi}.}
\label{fig:bias_SNR}
\end{figure}

Shear in each field was estimated as a weighted average of galaxy ellipticities using approximate inverse-variance weights, with unresolved galaxies assigned zero weight. Uncertainties were evaluated via 1000 bootstrap resamples of the galaxy populations. Multiplicative and additive shear biases were inferred from the best-fit linear relation between measured and input shears. To explore the dependence of bias on SNR, shear was measured from simulated populations with the same parameter distributions but varying lower flux limits. The resulting bias estimates for galaxy populations with $\textrm{SNR} > 25$ (corresponding to $I_0 \geq 50~\mu$Jy, i.e., the same flux range as in the image-domain analysis) are:
\begin{equation*}
\begin{aligned}
    \hat{M}_1&=(28.1\pm9.8)\times10^{-3},~\hat{C}_1=(6.5 \pm 2.6)\times10^{-4}; \\
    \hat{M}_2&=(31.8\pm9.8)\times10^{-3},~\hat{C}_2=(5.4 \pm 2.6)\times10^{-4}.
\end{aligned}
\end{equation*}

Similar to the image-domain analysis, shear estimates improve significantly with increasing SNR, with biases approaching the SKA-Mid requirements for high-SNR galaxies. This behavior is illustrated in Fig.~\ref{fig:bias_SNR}.

\subsection{Toward Robust Shear Estimation in Radio}
While both image- and visibility-domain analyses achieve additive biases close to the SKA-Mid requirements \citep{brown2015}, the multiplicative bias exceeds it by roughly an order of magnitude. Applying quality cuts based on galaxy properties or SNR can mitigate this bias to some extent, but at the cost of discarding a substantial fraction of usable galaxies. The high multiplicative bias likely reflects limitations of the shear estimator itself rather than the underlying shape measurements.

This challenge is well known in optical studies, where shear biases often surpass cosmological requirements \citep{Noise_bias}. A standard mitigation strategy is to average over a large galaxy sample~\citep{bias_req}, but this is less feasible for SKA-Mid due to its lower source density compared to contemporary optical surveys. Another common approach is post-measurement bias calibration~\citep{Conti_2017,Metacal}, where calibration parameters are derived from simulations. However, adapting such methods to radio datasets is challenging because of the high computational cost of visibility-domain simulations. Bayesian model-fitting techniques~\citep{Noise_bias_3, Noise_bias_2} have also been explored extensively, and more recently, deep learning–based shear estimators have shown promise~\citep{Shear_Measurement_ML,Forklens}. These emerging shape measurement methods, combined with appropriately tailored calibration strategies, represent possible avenues for improving shear estimation in radio surveys, but require dedicated testing on SKA-like data.
\section{Source Separation}
In optical imaging, overlapping sources can hinder accurate shape measurement, since most algorithms rely on extracting clean, isolated galaxies. This is particularly problematic for deep, wide-field surveys, where high source densities increase the likelihood of blending. While optical blending is already a significant challenge, radio interferometric imaging introduces additional complexities.

Unlike optical images—where sources appear as localized peaks in pixel space—radio interferometers record data as visibilities, which represent the spatial coherence of the sky emission integrated across the primary beam. Each visibility measurement contains contributions from all sources within the field, so information about individual objects is inherently mixed in the Fourier domain, with no single visibility corresponding to a unique sky position. The incomplete sampling of the Fourier plane further complicates image reconstruction, producing point spread functions (PSFs) with extended sidelobes. As a result, bright sources—even those far from the target—can generate noise-like artifacts in dirty images, obscuring fainter objects.

While these effects can be mitigated through full deconvolution using major–minor imaging cycles~\citep{CASA}, such methods are computationally expensive. Additional contamination arises from bright sources outside the main field of view, which can appear through the secondary sidelobes of the primary beam~\citep{FSCN}. These sidelobe effects are particularly problematic at low radio frequencies (e.g., SKA-Low), though their impact on SKA-Mid observations remains uncertain. Techniques such as peeling~\citep{Peeling} can help suppress these artifacts but require precise positional and morphological information for the brightest sources.

Addressing these challenges requires modeling multiple sources jointly in the visibility domain \citep{Chang_2004, Rivi3, Malyali}. While effective, this approach can be computationally intensive and depends on prior knowledge of source positions. An alternative is to develop algorithms that can disentangle overlapping emissions while preserving the subtle morphological features essential for weak-lensing analyses. In the optical domain, a range of analytical~\citep{sextractor,SCARLET} and machine-learning approaches \citep{GAN_deblender,Arcelin_2020} have emerged for this task. By contrast, radio-based source separation methods are still relatively limited, with most focusing on galaxies of simple morphology~\citep[see][and references therein]{Radio_object_ml}. With targeted development and validation, next-generation source-separation methods for radio could overcome current limitations and enable the high-precision shape measurements required for forthcoming radio weak-lensing surveys.
\section{Conclusions}
In this chapter, we evaluated the feasibility of conducting weak lensing experiments with SKA-Mid, with a particular emphasis on the challenges of practical shape measurement. To this end, we developed a dedicated simulation pipeline designed to generate realistic radio interferometric datasets, including instrumental effects, visibility sampling, noise, and image reconstruction artifacts, to provide a controlled testbed for evaluating shear measurement techniques in the radio domain.

Using this framework, we benchmark three representative shape measurement methods that have been proposed for radio weak lensing: SuperCALS, bias calibration in the image domain; RadioLensfit, a parametric model fitting technique in the visibility domain; and DeepShape, a deep learning-based estimator. All three methods demonstrate the capability to recover galaxy shapes with linear bias levels that are broadly comparable to, though not yet fully consistent with, the stringent SKA-Mid requirements \citep{ska_wl_ch_patel}. This confirms the progress made in adapting both traditional and modern approaches for radio data, while also highlighting the need for further refinement.  Shear estimates from simulated data showed additive biases consistent with SKA-Mid weak lensing targets \citep{brown2015}, but multiplicative biases an order of magnitude too high. This suggests that future radio weak lensing studies will likely require either more sophisticated shear estimators or an additional calibration step to achieve the required precision.

While not included in this analysis, source separation is a key obstacle to shear estimation with radio data. In radio surveys, incomplete Fourier sampling and complex sidelobes make disentangling galaxies more difficult than in the optical. While traditional deconvolution and peeling exist, machine-learning–based disentanglement methods are emerging as promising solutions. Addressing this issue will be essential for realizing weak lensing science with the SKA-Mid.
\section*{Author Ordering}
Authors for this chapter are ordered according to their overall level of contribution, in line with that expected for a small author list publication.

\section*{Acknowledgments}
{\small PT is funded by the National Research Agency (ANR) under the project ANR-22-CE31-0014-01 TOSCA.}

\bibliographystyle{abbrvnat}
\bibliography{chapter} 
\end{document}